\documentclass[preprint2,tighten,usenames,dvipsnames,breaklinks,colorlinks,citecolor=blue]{aastex6}
\pdfoutput=1 
\usepackage{amsmath,amstext}
\usepackage[T1]{fontenc}
\usepackage{apjfonts} 
\usepackage[figure,figure*]{hypcap}
\usepackage{amssymb}	
\usepackage{epsf}
\usepackage{epstopdf}
\usepackage{xspace}
\usepackage{amsmath}
\usepackage{xcolor}
\usepackage{hyperref}
\usepackage{subfigure}

\def\stacksymbols #1#2#3#4{\def\theguybelow{#2}
        \def\verticalposition{\lower#3pt}
        \def\spacingwithinsymbol{\baselineskip0pt\lineskip#4pt}
        \mathrel{\mathpalette\intermediary#1}}
\def\intermediary #1#2{\verticalposition\vbox{\spacingwithinsymbol
        \everycr={}\tabskip0pt
        \halign{$\mathsurround0pt#1\hfil##\hfil$\crcr#2\crcr
                \theguybelow\crcr}}}

\def\lta{\lesssim}
\def\gta{\gtrsim}

\def\Msun{M_\odot}

\def\kpc{\rm kpc}

\def\FeXXV{\ion{Fe}{25} }
\def\kms{{\rm km\:s^{-1}}}

\newcommand{\be}{\begin{equation}}
\newcommand{\ee}{\end{equation}}
\newcommand{\bea}{\begin{eqnarray}}
\newcommand{\eea}{\end{eqnarray}}

\shorttitle{Gas motions in cluster cores}
\shortauthors{Lau et~al}
\slugcomment{Submitted to the Astrophysical Journal}
\begin{document}

\title{Physical Origins of Gas Motions in Galaxy Cluster Cores: \\
Interpreting {\em Hitomi} Observations of the Perseus Cluster}

\author{Erwin T.~Lau\altaffilmark{1,2}}
\author{Massimo Gaspari\altaffilmark{3,5}}
\author{Daisuke Nagai\altaffilmark{1,2,4}}
\and
\author{Paolo Coppi\altaffilmark{1,2,4}}

\altaffiltext{1}{Department of Physics, Yale University, New Haven, CT 06520, USA; erwin.lau@yale.edu}
\altaffiltext{2}{Yale Center for Astronomy and Astrophysics, Yale University, New Haven, CT 06520, USA}
\altaffiltext{3}{Department of Astrophysical Sciences, Princeton University, 4 Ivy Lane, Princeton, NJ 08544-1001 USA}
\altaffiltext{4}{Department of Astronomy, Yale University, New Haven, CT 06520, USA}
\altaffiltext{5}{{\em Einstein} and {\em Spitzer} Fellow}

\begin{abstract}
The {\em Hitomi} X-ray satellite has provided the first direct measurements of the plasma velocity dispersion in a galaxy cluster.  It finds a relatively ``quiescent' gas with a line-of-sight velocity dispersion $\sigma_{v, \rm{los}}\simeq 160\;\kms$, at $30-60$ kpc from the cluster center. This is surprising given the presence of jets and X-ray cavities that indicates on-going activity and feedback from the active galactic nucleus (AGN) at the cluster center.  Using a set of mock {\em Hitomi} observations generated from a suite of state-of-the-art cosmological cluster simulations, and an isolated but higher resolution simulation of gas physics in the cluster core, including the effects of cooling and AGN feedback,  we examine the likelihood of {\em Hitomi} detecting a cluster with the observed velocities. As long as the Perseus has not  experienced a major merger in the last few gigayears, and AGN feedback is operating in a ``gentle" mode, we reproduce the  level of gas motions observed by {\em Hitomi}.  The frequent mechanical AGN feedback generates net line-of-sight velocity dispersions $\sim 100-200\;\kms$, bracketing the values measured in the Perseus core.  The large-scale velocity shear observed across the core, on the other hand,  is generated mainly by cosmic accretion such as mergers.  We discuss the implications of these results for AGN feedback physics and cluster cosmology and progress that needs to be made in both simulations and observations, including a {\em Hitomi} re-flight and calorimeter-based instruments with higher spatial resolution.
\end{abstract}



\section{Introduction}
\label{sec:intro}


Gas motions within clusters of galaxies provide important clues about the astrophysics and the formation of galaxy clusters, as the level of gas motions are influenced by baryonic physics and the cluster accretion history. The feedback from the AGN in the center of massive cluster galaxies, has been identified as a potential source for injecting energy into the intracluster medium (ICM) gas 
\cite[e.g.,][]{churazov_etal02, McNamara:2007, Gaspari:2012b}. 
The amount of energy provided by the AGN can help offset radiative cooling occurring in cool-core clusters where the cooling time is short compared with the age of the universe (typically within $r\lta100$\,kpc) yet evidence for large amounts of cooled gas has not been found.  Bipolar AGN jets/outflows are likely to impart the kinetic energy to the ICM by converting into heat through the diffusion and dissipation of turbulence stimulated by the jets and outflows, alongside weak shocks and buoyant hot bubbles.

At the same time, cosmic gas accretion, such as mergers, infalling galaxies and groups, and accretion from filaments, can also drive gas motions and transfer energy to the ICM \citep[e.g.,][]{norman_bryan99, dolag_etal05, vazza_etal11, nelson_etal12,schmidt_etal14,miniati15}. These accretion processes stir up the hot gas in the gravitational potential well of the cluster, generating bulk and turbulent motions which provide additional pressure support \citep[e.g.,][]{rasia_etal04, lau_etal09, shi_etal14} and introduce a bias in the hydrostatic equilibrium (HSE) mass estimates of galaxy clusters 
\citep[e.g.,][]{rasia_etal06, nagai_etal07b, nelson_etal14, Biffi:2016, henson_etal17}. The hydrostatic mass bias is a major systematic uncertainty in current cluster-based cosmological constraints \citep{allen_etal08, vikhlinin_etal09, planckXXIV15}. The level of gas motions also impact astrophysical and cosmological constraints from Sunyaev-Zel'dovich \citep{shaw_etal10, battaglia_etal12, Khatri:2016} and radio measurements \citep[e.g.,][]{eckert_etal17}. Penetrating gas streams originating from large-scale filaments can reach the cluster core and dissipate their kinetic energies in the cluster core through weak shocks and turbulence, providing additional heat source to offset cooling \citep{zinger_etal16}.  

The energy contained in turbulence generated from AGN and cosmic accretion can be constrained via the relative fluctuations of the X-ray surface brightness \citep[e.g.,][]{Gaspari:2013_coma} as has been applied to 33 clusters with deep {\em Chandra} data \citep{Hofmann:2016} and the nearby X-ray bright Perseus cluster  \citep{zhuravleva_etal14b}. These estimates of turbulence via surface brightness fluctuations are indirect and can be somewhat model-dependent. Ideally, as in studies of the interstellar medium of the Milky Way Galaxy, we can constrain ICM turbulence more directly via the Doppler shifts and broadening of ICM emission lines \citep[e.g.,][]{inogamov_sunyaev03, bruggen_etal05, biffi_etal13}. Unfortunately, this is not easy at X-ray energies. During the past decade, only upper/lower limits on the level of turbulent motions have been available via X-ray grating spectroscopy measurements \citep[e.g.,][]{Sanders:2013,Ogorzalek:2017}. While these limits are consistent with those predicted from simulations \citep[e.g.][]{vazza_etal13}, more precise measurements are required. 

Recently, despite its tragically short lifespan, the {\em Hitomi} X-ray satellite \citep[formerly known as ASTRO-H,][]{takahashi_etal14} provided the first direct measurements of core gas motion \citep[hereafter H16]{hitomi16} via its Soft X-ray Spectrometer \citep[SXS][]{mitsuda_etal14}, an X-ray calorimeter which enables 
high-spectral resolution ($4.9$~eV) measurements of Doppler line shifts and broadening. The main result is that the line-of-sight velocity dispersion in the Perseus core is measured to be $164\pm 10\,\kms$ between cluster-centric radius of $R\simeq 30 - 60$~kpc ($187\pm13\,\kms$ within $30$~kpc). 
This level of gas motion has been dubbed unexpectedly low, as the presence of buoyant bubbles and surface brightness fluctuations suggests ongoing AGN feedback activity. This apparent tension in the Perseus cluster can potentially change our understanding of AGN feedback physics in galaxy clusters. 

In this paper, we explore the velocity constraints imposed by the apparently ``quiescent'' line-of-sight gas motions measured with {\em Hitomi}. 
We use a combination of simulations to study the physical origin of gas motions in galaxy cluster cores, covering physical processes on both large and small scales.  In order to allow a proper comparison with {\em Hitomi} data, mock {\em Hitomi} observations are generated from the simulation data to explicitly account for effects such as the significant spatial averaging that results from the relatively low angular resolution of {\em Hitomi}. 

On larger radial scales, to quantify the effects of cosmic accretion and the inhomogeneous density and velocity structures that it produces, we study the level of core gas motions in a mass-limited sample of galaxy clusters extracted from large cosmological hydrodynamical simulation {\em Omega500} \citep{nelson_etal14} that self-consistently follow cosmic accretion and mergers. 

In agreement with prior conclusions based on the morphology of the Perseus cluster emission, we find that the core region must be dynamically relaxed. In particular, there must not have been a major cluster merger in the last few gigayears.  Ignoring the effects of baryonic physics such as radiative gas cooling and AGN feedback, ``relaxed" clusters with core gas velocity dispersions $\sim 100-200 \kms$ are not rare in our simulations. From this restricted point of view, the {\em Hitomi} result is not that unusual. 

As is well-known, however, baryonic physics cannot be ignored in the core of Perseus. Turning on radiative cooling and thermal AGN feedback in cosmological simulations, we do find a possible tension with observations. Even for relaxed clusters, the predicted gas velocity dispersions are typically larger than the {\em Hitomi} results. This, however, is likely due to the relatively simplistic subgrid thermal AGN feedback model employed in cosmological simulations which typically do not have the spatial resolution to resolve AGN feedback physics. Subgrid AGN feedback models such as ours are known to have difficulties matching the thermodynamic properties of cluster cores. 

High resolution isolated cluster simulations with ``gentle'' kinetic feedback model, on the other hand, are more successful in explaining the thermodynamic properties in relaxed clusters \citep[from][G12]{Gaspari:2012a}. These simulations are among the most detailed studies yet of AGN self-regulation in a cluster environment (despite exclusion of potentially important physics such as magnetic fields). We find that the level of velocity dispersion in such simulations is consistent with the {\em Hitomi} observations. In quasi-steady state (where gas accretes onto the central blackhole fairly regularly and the AGN undergoes frequent small outbursts), the AGN-driven gas motions reasonably bracket the ``quiescent" level seen by {\em Hitomi}. However, a large-scale velocity shear observed across the {\em Hitomi} observation field, is unlikely to be explained by AGN feedback alone as the velocity fields from AGN-induced turbulence are rather stochastic. On the other hand, velocity flows associated with cosmic accretion do not have these problems and can explain the level of shear observed by {\em Hitomi}. It thus appears that the level of gas motions observed by {\em Hitomi} in the core of Perseus are in fact {\em consistent} with the expectations of cosmic accretion {\em combined} with gentle AGN feedback. 
In this interpretation, then, the core of Perseus is ``quiescent" and relaxed in the sense that nothing dramatic has happened to it in the last gigayear or so, but it is {\em not} quiet.    

The paper is organized as follows. Section~\ref{sec:methods} describes our suite of numerical simulations (both cosmological and isolated clusters) and the synthetic {\em Hitomi} observations made from them.  Results on the predicted velocity dispersions and bulk motions are presented and analyzed in Section~\ref{sec:results}. We discuss their implications for AGN feedback and cluster astrophysics and cosmology in Section~\ref{sec:disc}.
We summarize the main conclusions, highlight the study limitations, and remark on future prospects in Section~\ref{sec:summary}.


\section{Simulations}
\label{sec:methods}


\subsection{Cosmological Simulations}

We use galaxy clusters extracted from the {\em Omega500} simulation \citep{nelson_etal14}, a cosmological hydrodynamical simulation performed with the Adaptive Refinement Tree code \citep{kra99, kra02, rudd_etal08}. The cosmology adopted is the flat $\Lambda$CDM model with WMAP five-year ({\em WMAP5}) cosmological parameters: $\Omega_m = 1 - \Omega_{\Lambda} = 0.27$, $\Omega_b = 0.0469$, $h = 0.7$ and $\sigma_8 = 0.82$, where the Hubble constant is defined as $100h$~km~s$^{-1}$~Mpc$^{-1}$ and $\sigma_8$ is the mass variance within spheres of radius 8$h^{-1}$~Mpc.  The simulation box has a comoving length of 500~$h^{-1}\, {\rm Mpc}$, resolved using a uniform $512^3$ root grid and 8 levels of mesh refinement, with maximum comoving spatial resolution of 3.8~$h^{-1}\, {\rm kpc}$.  
    
In this paper, we analyze a mass-limited sample of 63 galaxy clusters with $M_{\rm 500c}\geq 3\times10^{14}h^{-1}M_{\odot}$ at $z=0.0$.
These clusters have been initially identified using a spherical overdensity halo-finder described in \citet{nelson_etal14}.  The final cluster sample is from a re-simulated box with higher resolution dark matter particles in regions of the identified clusters.  This ``zoom-in'' technique results in an effective number of particles of $2048^{3}$, corresponding to a dark matter particle mass of $9\times10^{8}\, h^{-1} M_{\odot}$ inside spherical region with cluster-centric radius of three times the virial radius for each cluster. 

To investigate the effect of AGN feedback on gas motions in the cosmological run, we re-simulate the same galaxy cluster halos with and without baryonic physics. The run without baryonic physics model is referred to as the non-radiative (NR) run, and the one with baryonic physics is referred to as the AGN run. Specifically, in the AGN run,  in addition to metallicity-dependent radiative cooling, star formation, and thermal feedback from supernovae,  we implement subgrid thermal AGN feedback, following the accretion and mergers of BH in dark matter halos as in \citet{booth_etal09}.  Briefly, BH particles are seeded with an initial mass of $10^5h^{-1}M_\odot$ at the centers of dark matter halos with $M_{500c} > 2 \times 10^{11}h^{-1}M_\odot$. Throughout cosmic history, these BH particles accrete gas with a rate given by a modified Bondi accretion model with a boost parameter of $\alpha = 100$\footnote{The boost 
parameter is introduced to artificially boost the Bondi accretion rate in order to compensate for the low BH accretion rate due to the lack of resolution in cosmological simulations \citep[see][for more detail]{booth_etal09}. Such boosting also compensates for the inadequacy of Bondi formula to track cold flows (\citealt{Gaspari:2013_CCA}).} 
and return the feedback energy as a fraction of the accreted rest mass energy ($\epsilon = 0.2$) into the environment in the form of thermal energy.  To keep the thermal feedback from the BH from immediately radiating away, following \citet{booth_etal09} we impose a minimum heating temperature, $T_{\rm min} = 10^7$~K, and require that the BH stores enough feedback energy until it accumulates enough energy to heat neighboring gas cells, each by an amount of $T_{\rm min}$. 

\subsection{High-Resolution Non-Cosmological Simulation of Self-Regulated AGN feedback}\label{sec:G12}

To understand in more detail the impact of AGN feedback in driving gas motions, 
we analyze high-resolution hydrodynamical simulations of an isolated cluster \citep[from][G12]{Gaspari:2012a} with self-regulation that has been proven to solve several aspects of the cooling flow problem \citep[for a brief review]{Gaspari:2015_xspec}. While cosmological simulations are excellent tools to study cosmic flows, they have limited resolution in space and time. The G12 simulations use over 20 times higher resolution to dissect the impact of AGN outflows/jets in the radiatively cooling ICM of cluster cores.
Below, we briefly summarize the main features and physical mechanisms. We refer the interested reader to G12 for more detailed information about the simulations. 

A typical massive, cool-core galaxy cluster is initialized in hydrostatic equilibrium in a static NFW halo ($M_{\rm vir}\sim 10^{15}\ \Msun$ and gas fraction 0.15). To mimic the effects of cosmic accretion on the ICM profiles (aka cosmic weather) and to avoid idealized initial conditions, the gas density and temperature profiles are initially perturbed by noise with $0.3$ relative amplitude, which subsequently induces subsonic perturbations in the velocity field of the order of $100\;\kms$ -- mimicking the chaotic motions driven in the large-scale cosmological runs.
The inner $500$~pc zones (the sink region) track the inflow of matter toward the central supermassive black holes (SMBH), which result to be dominated by the accretion of cold gas. The developed phenomenon is also known as chaotic cold accretion (CCA -- studied in-depth in \citealt{Gaspari:2013_CCA}).  
In the CCA regime, the SMBH accretion is driven by the cold gas clumps condensed out from turbulence-induced thermal instability (TI). Subsonic chaotic motions induce the non-linear growth of thermal instability that causes warm filaments and cold clouds to condense out of the hot ICM phase, as the cooling time to free-fall time ratio is roughly $t_{\rm cool}/t_{\rm ff}\lesssim 10$. The condensed filaments and clouds rain onto the nuclear region, experience loss of angular momentum through recurrent stochastic inelastic collisions, and cause rapid boost in the SMBH accretion rate \citep{Gaspari:2017_CCA}.  
The present simulated cluster has minimum TI-ratio of 7 at $r\sim5\,$kpc as for a typical massive cool-core cluster, thus enabling to efficiently trace the fully developed CCA rain. Larger ratios, in particular above 20, would make the system a non-cool-core object not prone to thermal instability for several Gyr due to the large cooling time.


In the isolated G12 cluster simulation, the self-regulated AGN feedback is implemented as follows. The tracked cold mass flux passing through the inner spherical sink region sets the SMBH accretion rate, which results to be comparable to the residual plasma cooling rate within the 5 kpc core. 
The AGN triggers injecting mass and kinetic energy through the nozzle, internal boundaries with power given by a fraction of the accreted rest mass energy, $P_{\rm agn} \equiv (1/2)\,\dot M_{\rm out} v^2_{\rm out} = \epsilon_{\rm BH}\,\dot M_{\rm sink}c^2$. The adopted macro kinetic efficiency is $\epsilon_{\rm BH}\approx5\times10^{-3}$. Increasing the efficiency above this value would overheat the cluster, while lowering it would allow a cooling flow catastrophe to develop -- both ruled out by observations (more below).
Typical velocities and mass outflow rates within the kpc scale are several $10^3\,\kms$ and several 10\,$\Msun$\,yr$^{-1}$, respectively, as the outflows entrain the surrounding mass \citep{Gaspari:2017_uni}. The outflows dissipate the AGN energy in the central $\sim 100$~kpc region initially via the cocoon shock, later with mixing buoyant bubbles, and finally with turbulence which isotropizes most of the injected energy.

The G12 runs (carried out using the Eulerian FLASH4 code) pay particular attention to numerical accuracy, by implementing an exact solver for the radiative cooling of the ICM plasma and using a third-order accurate PPM solver for the hydro part, together with conservative timestep limiters and multiple static nested meshes which avoid derefinement noise.

The G12 simulations have been shown to solve several long-standing issues in the cooling flow problem \citep{Gaspari:2013_rev} and are in good agreement with {\em Chandra} and {\em XMM} data \citep{McNamara:2012}. In particular, the
recurrent, self-regulated and gentle AGN outflows seen in the simulations are able to quench the cooling of the core gas to net rates that are $5\%-10\%$ those inferred from its X-ray luminosity, while at the same time preserving the cool-core structure for several Gyr -- along with the positive temperature gradient and the mild inner density variations that are usually seen. Moreover, the emission measure of the X-ray spectrum is properly suppressed by 2 orders of magnitude from the hard to soft X-ray, as constrained by {\em XMM}-RGS data \citep[e.g.,][]{Peterson:2003}. The simulations reproduce the major imprints of AGN feedback, such as under-dense X-ray cavities, weak shocks, and metal outflows up to $\sim 100$~kpc.
Finally, the CCA feeding is linked to the formation of extended warm filaments and cold clouds which result from the multiphase condensation cascade. These gaseous structures are co-spatial with the soft X-ray gas, as found by multi-wavelength observations of ionized, neutral, and molecular gas \citep[e.g.,][]{McDonald:2011a,Werner:2014,Tremblay:2016}.

\section{Analysis methods}

\begin{figure}[t]
\begin{center}
\includegraphics[scale=0.35]{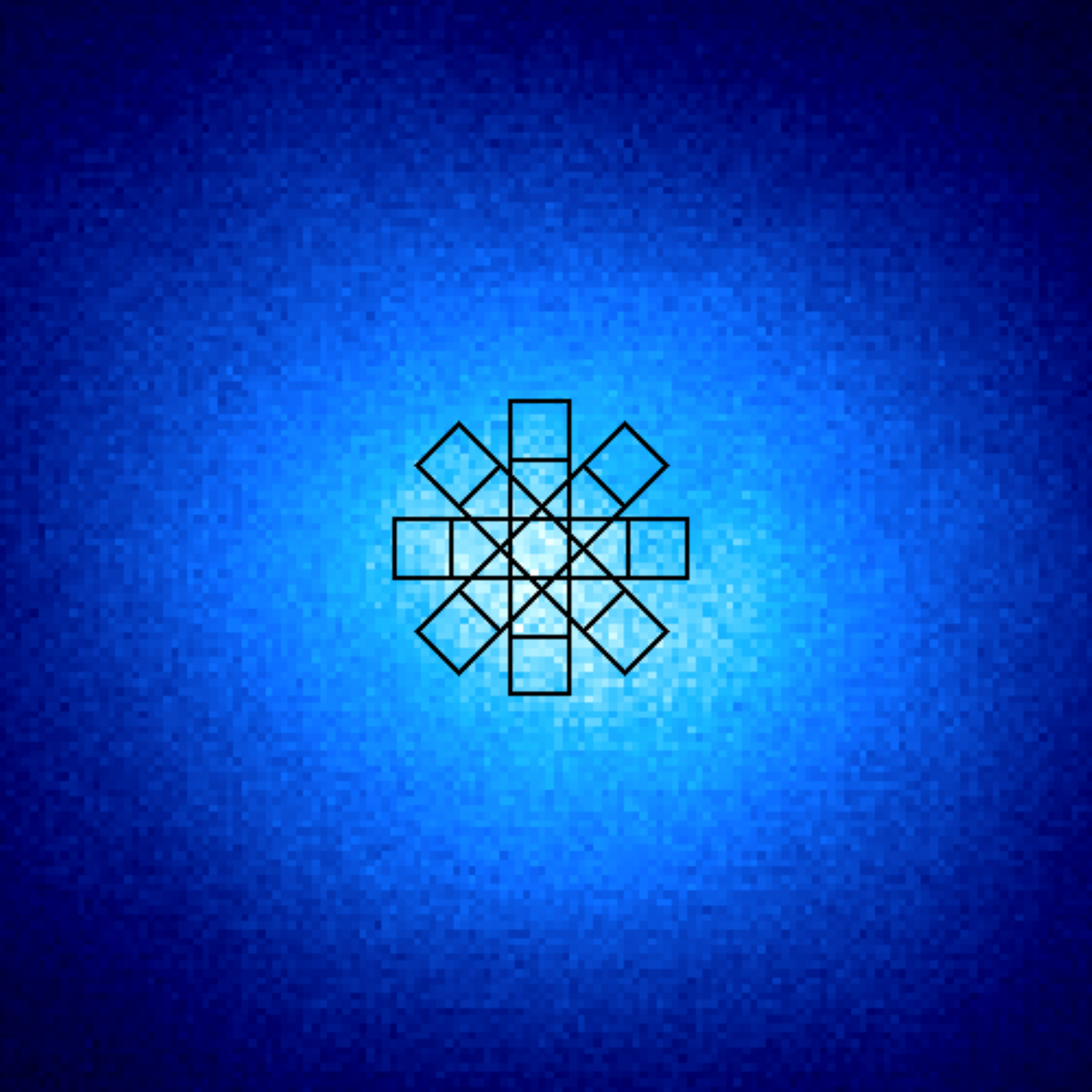}
\caption{
Mock {\em Hitomi} map for one of the simulated clusters in the {\em Omega500} sample, showing the core region. The width of the
region is 550~kpc. The squares,
each with size $30\,{\rm kpc}\times 30\,{\rm kpc}$, indicate the regions where we extract the spectra.  
}
\label{fig:region}
\end{center}
\end{figure}

In this section, we describe how we measure the ICM velocities from our cosmological and isolated simulations, for comparison with the {\em Hitomi} measurements. For the cosmological simulations, we derive bulk and turbulent velocities by fitting the mock {\em Hitomi} spectra extracted from the simulations. For the isolated simulations with over 500 snapshots, we adopt a simpler approach where the measured velocities are weighted by hard X-ray emission as described below.

\subsection{Cosmological Simulations}

We generate mock {\em Hitomi} photon maps and
spectra to measure the level of gas motions in the simulated cosmological clusters, using the same pipeline used in \citet{nagai_etal13}. 
For each cluster, we first generate the flux map from the simulation output. 
For each hydrodynamic cell within a cubic volume of length $5h^{-1}$~Mpc centered on the cluster, 
we compute the emissivity $\epsilon_E$ for each energy $E \in [6.5,7.5]$~keV 
with energy bin size of $\Delta E=1$~eV, using the APEC plasma code \citep{smith_etal01} 
with AtomDB version 2.0.2 \citep{foster_etal12}.  
The emissivity $\epsilon_E = \epsilon_E(\rho,T,Z,z_{\rm obs}, v_{\rm LOS})$ 
is a function of the gas density $\rho$, temperature $T$, the gas metallicity $Z=0.3Z_\odot$\footnote{ This value of the metallicity is motivated by the observational results in \citet{werner_etal13} which shows that the metallicity of the ICM in Perseus is close to $0.3\,Z_\odot$ uniformly throughout the cluster. }, 
and line-of-sight velocity $v_{\rm LOS}$ of the cell. 

We include thermal broadening in the emission lines. 
The observed redshift of the clusters is set to $z_{\rm obs}=0.01756$, the redshift of the Perseus Cluster.  

Each flux map is then convolved with the {\em Hitomi} ARF ({\tt sxt-s\_140505\_ts02um\_intall.arf}) and RMF ({\tt ah\_sxs\_5ev\_20130806.rmf}) response files from SIMX\footnote{\url{http://hea-www.harvard.edu/simx/}}. 
The energy resolution of the RMF file is $5$~eV. Photons for each location are then drawn from the convolved flux map assuming a Poisson distribution.  We assume a galactic column density of $N_H = 2\times10^{20}~{\rm cm^{-2}}$. 
Changing the column density value to other values, such as the estimated value for Perseus $N_H = 1.3\times10^{21}~{\rm cm^{-2}}$ has no significant effect on our results. 
To ensure our mock simulated spectrum is not limited by statistical uncertainties,
we set the exposure time to be $t_{\rm exp}= 2$~Ms per pointing to ensure enough photons in the spectrum. 
We do not model background noise because it is sub-dominant to the strong \FeXXV~K$\alpha$ line complex where the gas velocity constraints come from. 

For each mock X-ray map, we extract spectra from regions laid out in 8 azimuthal arms. 
Each arm contains three spectral regions, with size $30\;{\kpc} \times 30\;{\kpc}$ each,  extending radially from the cluster center (see Figure~\ref{fig:region}). These roughly correspond to the regions of the Perseus Cluster measured by {\em Hitomi} (H16).
We select radial distance from the cluster center of $R = [30, 60]$~kpc as in the main {\em Hitomi} observation. 
For each spectral region, we measure both bulk velocity and velocity dispersion of gas by performing spectral fitting for each spectrum. The spectral fitting is performed by using XSPEC version 12.8. The spectral model consists of the BAPEC model and the Galactic absorption model \citep[the {\tt wabs} model by][]{morrison83} and is convolved by the detector and telescope responses. The gas temperature, redshift, velocity dispersion, and the normalization are allowed to vary while the hydrogen column density and metal abundance are fixed at $N_H = 2\times10^{20}~{\rm cm^{-2}}$ and $Z=0.3\,Z_\odot$ respectively, the values adopted when generating the mock simulations. For metal abundance, the tables in \citet{anders89} is used. The velocity dispersion given by the Gaussian $\sigma$ returned by the BAPEC model fit. The bulk velocity along the line of sight is calculated by $c(z - z_{\rm cluster})$ where $z$ is the fitted redshift and $z_{\rm cluster} = 0.01756$. 

\subsection{Isolated AGN Feedback Simulation}

Given that the G12 simulations have a very large number of high-resolution snapshots, 
it is not feasible to run the full machinery described above. The LOS variability seen 
in the simulations (due to which direction one chooses to observe the cluster from) also swamps any instrumental uncertainty. We thus opted for a faster, yet still realistic, synthetic approach. The kinematic properties (such as bulk velocities and dispersions) are retrieved by weighting the LOS velocity by the luminosity\footnote{Plasma luminosity in a zone with volume $dV$ is $dL=n_{\rm e}\, n_{\rm i}\,\Lambda\,dV$, where $n_{\rm e}$, $n_{\rm i}$ are the electron and ion number density, and $\Lambda(T,Z)$ is the plasma cooling function in collisional ionization equilibrium with metallicity $Z=0.3\,Z_\odot$ (\citealt{Sutherland:1993}).} of the \ion{Fe}{25}-\ion{Fe}{26} gas, which is mostly tied to the hard X-ray gas. 
Numerically, for each cylindrical region corresponding to the chosen projected aperture ($R = [30,60]$~kpc or $R < 30$~kpc) we compute the weighted mean/bulk velocity as
\begin{equation}
\bar{v} = \frac{\sum_j v_j \epsilon_j  \Delta V_j}{\sum_j \epsilon_j \Delta V_j}
\end{equation}
where $\epsilon_j = n_{e,j}\,n_{i,j} \Lambda(T_j ; Z=0.3\,Z_\odot )$ is the emissivity of each $j$-th gas cell with temperature $5 < T_j/{\rm keV} < 9$ along the line of sight, and $\Delta V_j$ is the related cell volume. The weighted velocity dispersion is similarly computed as
\begin{equation}
\sigma_v^2 = \frac{\sum_j (v_j-\bar{v})^2 \epsilon_j  \Delta V_j}{\sum_j \epsilon_j \Delta V_j}.
\end{equation}
The ICM velocities computed with this method are very close to those derived from the mock spectra. 
We then compute and store these kinematic quantities for several lines of sight, chosen to have random inclinations. While simple in nature, such an algorithm should capture the main features of the simulations that are relevant to the {\it Hitomi} observation, i.e., the properly averaged, high-resolution LOS kinematics of the hard X-ray emitting plasma. 


\section{Results}
\label{sec:results}


\begin{figure*}
\begin{center}
\subfigure{\includegraphics[scale=0.465]{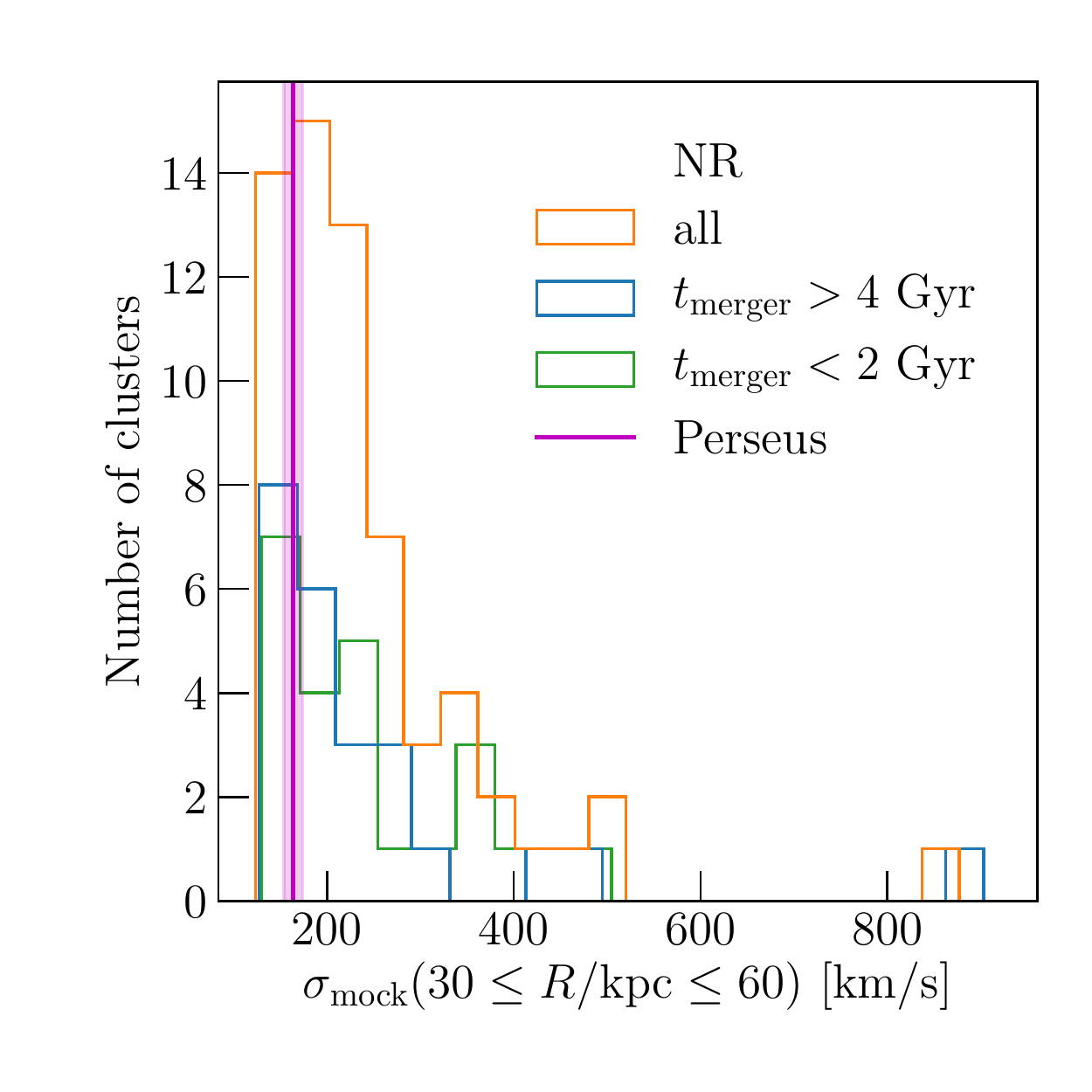}}
\subfigure{\includegraphics[scale=0.465]
{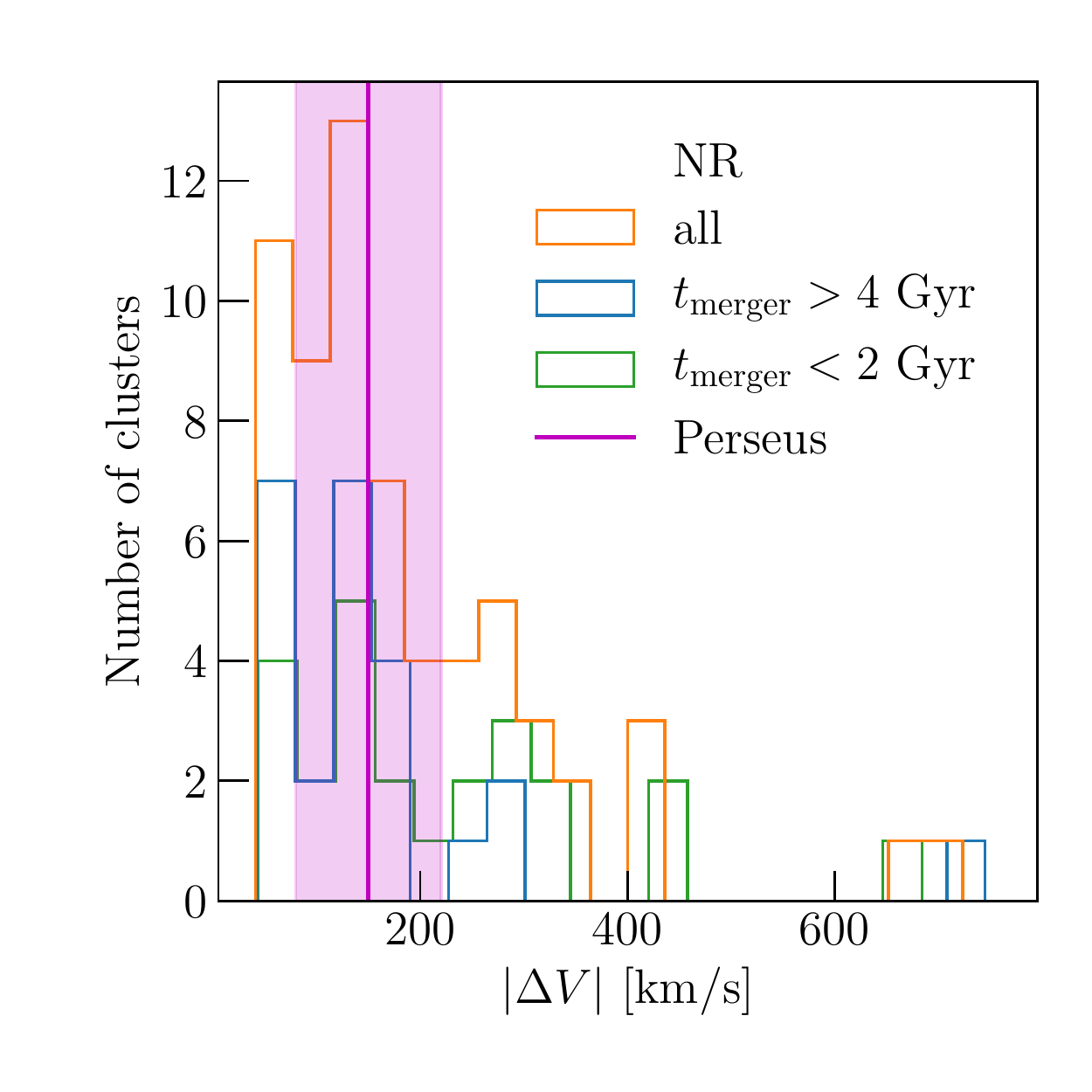}}
\subfigure{\includegraphics[scale=0.465]
{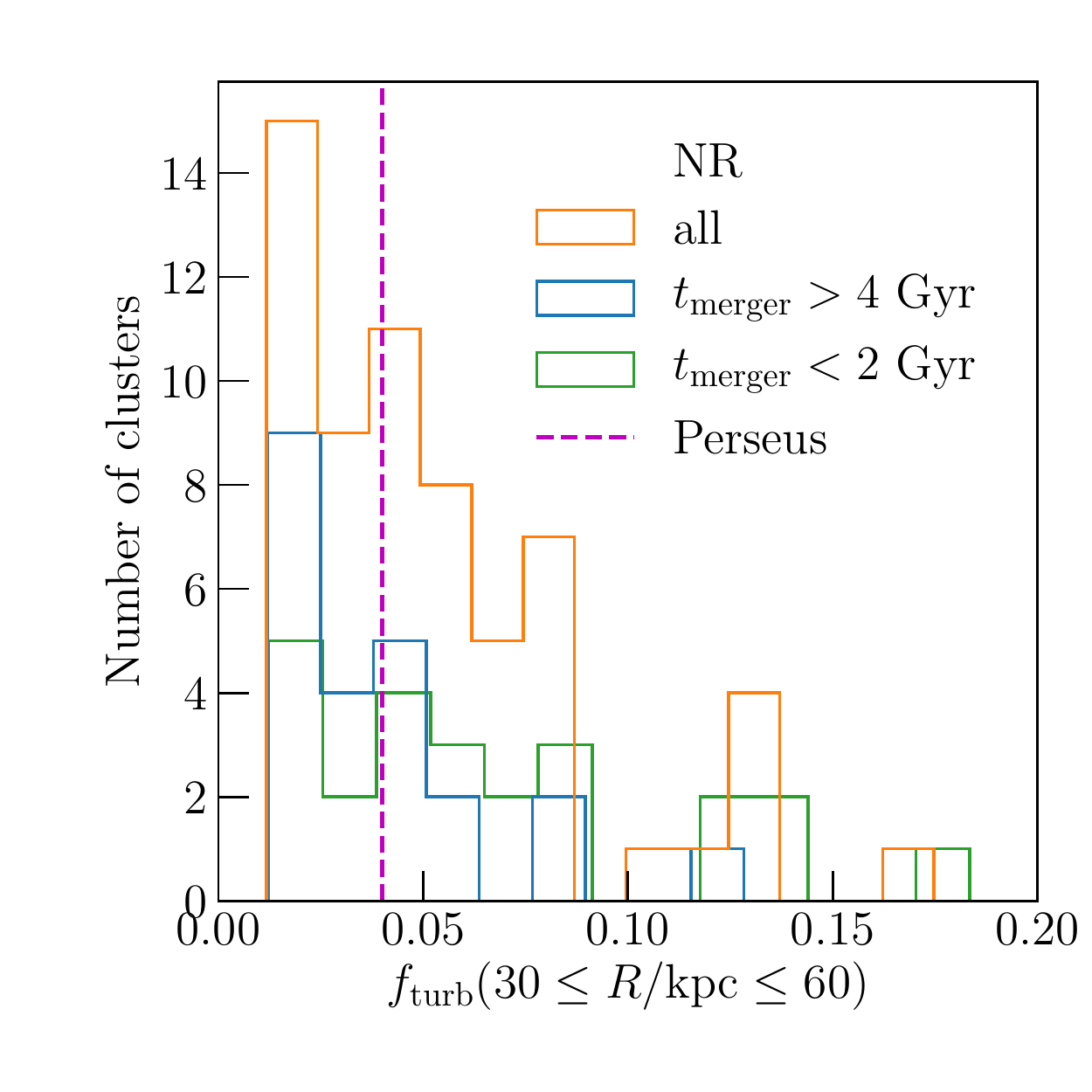}}
\subfigure{\includegraphics[scale=0.465]{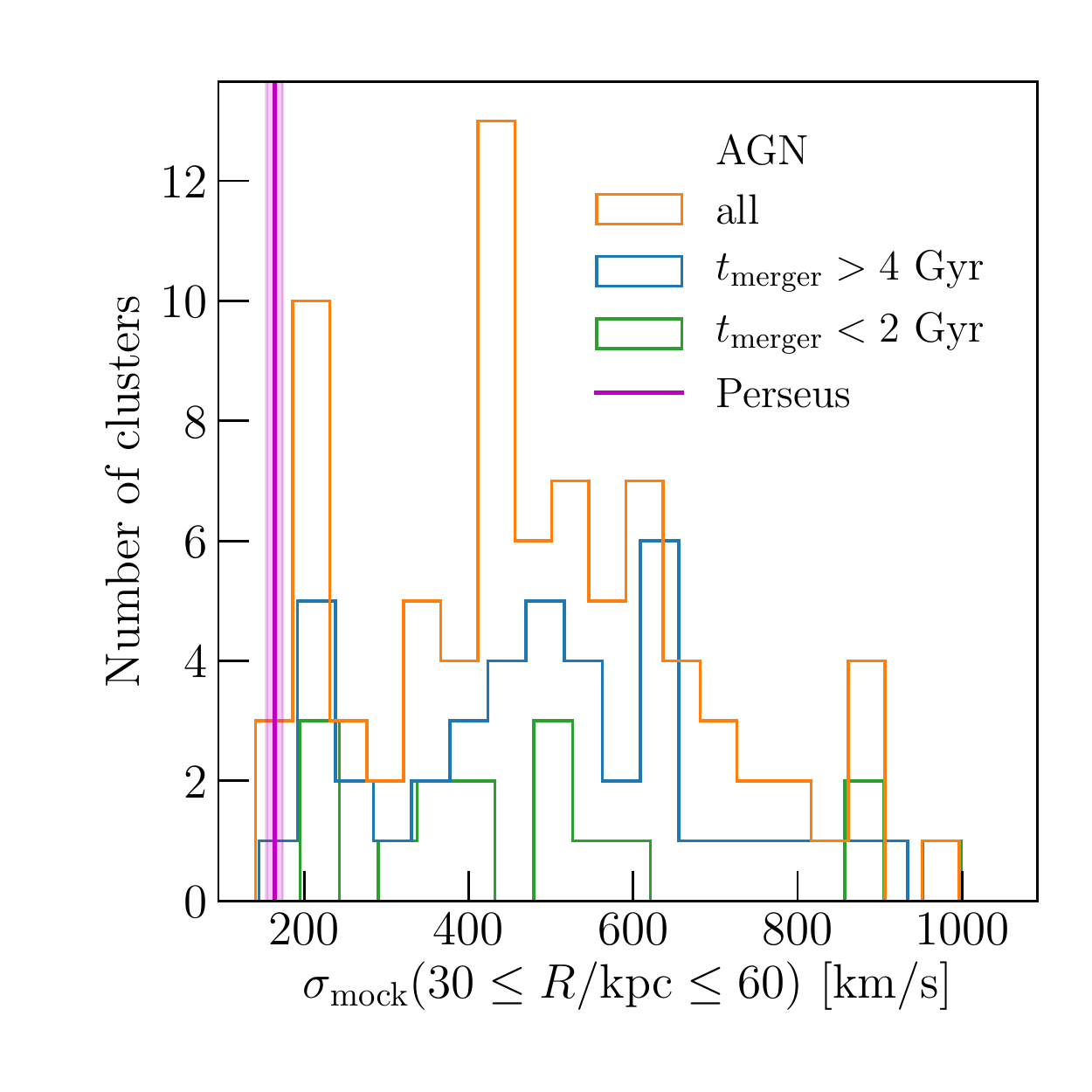}}
\subfigure{\includegraphics[scale=0.465]
{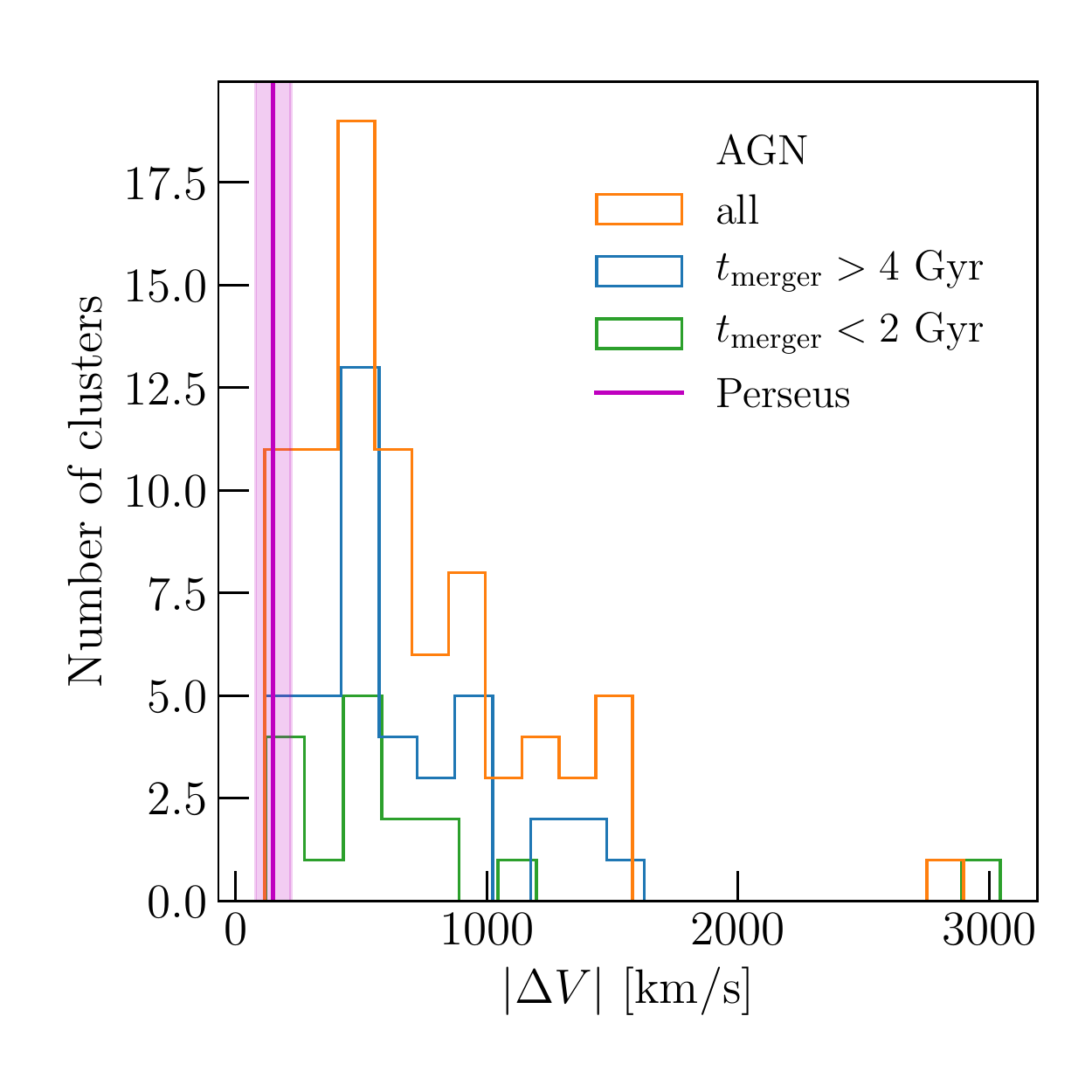}}
\subfigure{\includegraphics[scale=0.465]
{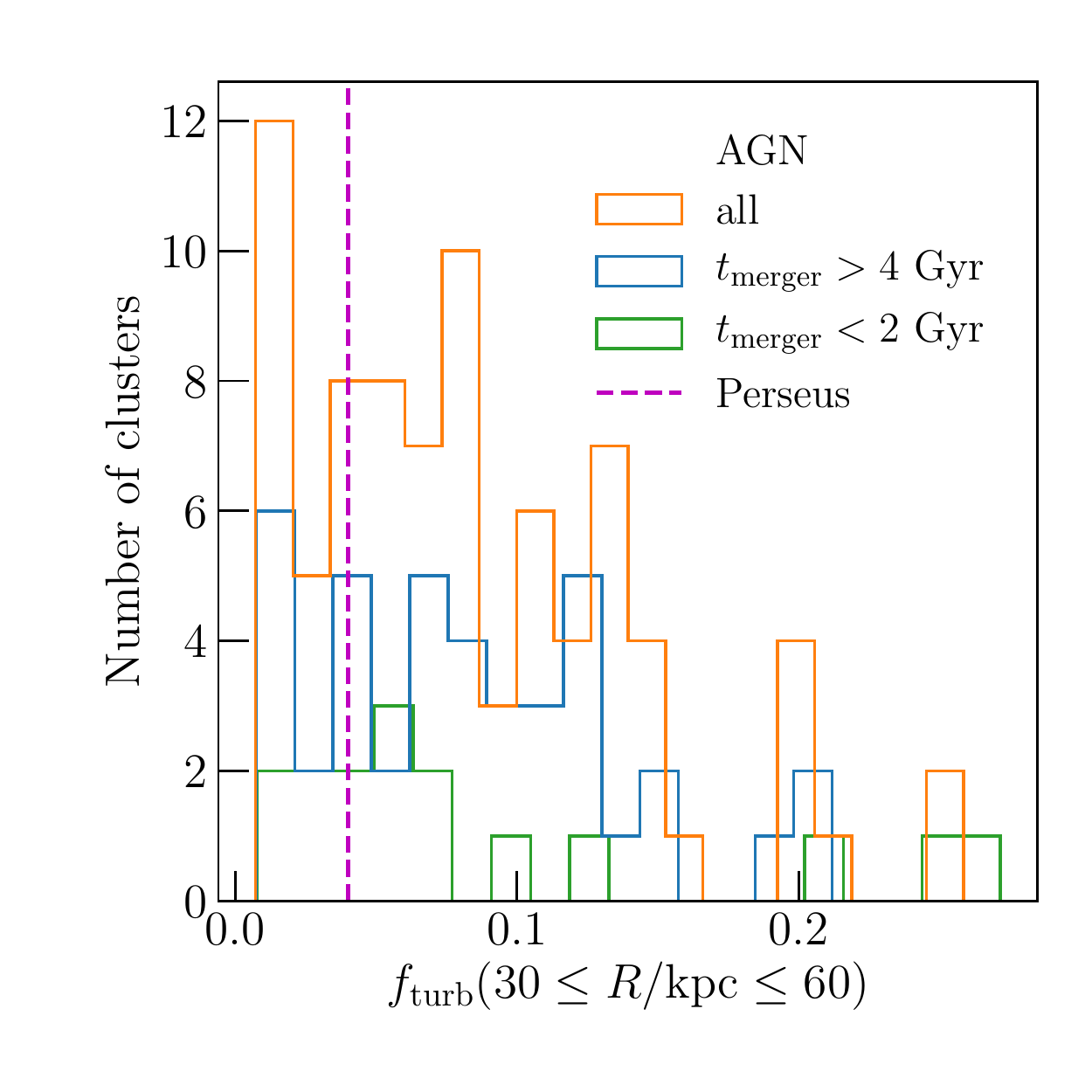}}
\caption{
Distributions of various velocity observables measured from mock {\em Hitomi} spectra for 63 clusters in the NR ({\em top} panels) and subgrid AGN ({\em bottom} panels) runs of the {\em Omega500} simulation.
{\em Left} panel: distribution of the line-of-sight (LOS) velocity dispersion measured from mock {\em Hitomi} spectrum for 63 clusters in the {\em Omega500} simulation measured within $30 \leq R/{\rm kpc} \leq 60$, as in H16. 
{\em Middle } panel: LOS velocity shear magnitude, defined as the difference between the bulk velocity in the inner radial bin $R < 30\,{\rm kpc}$ and the outer radial bin $30 \leq R/{\rm kpc} \leq 60$.
{\em Right} panel: the ratio of turbulent pressure to thermal pressure, computed as the LOS velocity dispersion squared divided by the  spectral temperature. 
The {\em orange}, {\em green} and {\em blue} lines represent the whole cluster sample, clusters with short and long $t_{\rm merger}$, the time since major merger,  respectively. The magenta lines indicate the {\em Hitomi} measurement for the Perseus Cluster, with the shaded regions indicating the $1\sigma$ error. 
}
\label{fig:cosmo}
\end{center}
\end{figure*}

\subsection{Results from Cosmological Simulation}
\label{sec:cosmo_sim}

\subsubsection{Cosmic-driven Gas Motions}
\label{sec:cosmo_sim_agn}

The top left panel in Figure~\ref{fig:cosmo} shows the distributions of the LOS velocity dispersion of the clusters in the NR run of the {\em Omega500} simulation, measured within the projected radial bin $R = [30, 60]$~kpc corresponding to the region observed in the Perseus cluster by {\em Hitomi}. The velocity dispersion is taken as the mean of the 8 arms. The NR clusters show a wide range of velocity dispersion, varying from $\sim 100\;\kms$ to $\sim 1000\;\kms$. Around a third of the clusters have velocity dispersion at around $200\;\kms$ where the distribution peaks. Incidentally, the value of the Perseus Cluster $\sim 160\,\kms$ is consistent with this peak value.  
To characterize the dependence of the level of core gas motions on the dynamical state of the cluster, we compute the time since major merger (with mass ratio $\leq 1/6$) for each clusters in the simulations \citep[see][for more details]{nelson_etal12}. Clusters that have not undergone a major merger for the past $4$~Gyr are `relaxed', while those have undergone a major merger in the past $2$~Gyr are `disturbed'. The relaxed clusters in our simulations in general have smaller velocity dispersion and velocity shear, with values closely matching to Perseus, which is a relaxed cluster based on its X-ray
morphology.

The top middle panel of Figure~\ref{fig:cosmo} shows the histogram of the velocity shear magnitude, $|\Delta V|$, of the NR clusters. The velocity shear is computed as the absolute difference in the bulk LOS velocity between the cluster center and $R=60$~kpc, taking the maximum of the 8 arms. While nearly a third of the NR clusters have small $|\Delta V| < 100\;\kms$,  there are considerable number of clusters (about a quarter of the total) with $|\Delta V|$ close to the Perseus value of $150 \pm 70\;\kms$. Having said that, careful interpretation should be given to the bulk LOS velocity from the {\em Hitomi} data, as it is nearly unconstrained in each observed pixel due to
the current systematic calibration error of $\pm\,50\;\kms$ (see more discussions in Section~\ref{sec:vbulk}).

The top right panel in Figure~\ref{fig:cosmo} shows the histogram of the turbulent-to-thermal pressure ratio $f_{\rm turb}$ in the same bin of $R = [30, 60]$~kpc. Following H16, we compute the turbulent pressure to thermal pressure ratio as 
\begin{equation}
f_{\rm turb} = \frac{\sigma_{\rm mock}^2}{kT_x/\mu m_p}
\end{equation}
where $T_x$ is the temperature measured from the X-ray spectrum, 
$\mu=0.59$ is the mean molecular weight and $m_p$ is the proton mass. 
The turbulent pressure fraction in our non-radiative clusters are small in the central region of the cluster. 
The distribution of the turbulent-to-thermal pressure ratio peaks at around $2\%$. Less relaxed clusters have higher turbulent pressure, with $f_{\rm turb} \approx 5\%$.  
The observed value for Perseus is $4\%$, which is only slightly higher than the averaged value from the cosmological simulation. 

\subsubsection{Cosmic-driven Gas Motions with subgrid thermal AGN Feedback}
\label{sec:cosmo_sim_agn}

While the NR simulation provides a {\em lower} limit to the level of gas motions, driven by cosmic accretion process such at mergers, we expect that the introduction of AGN feedback and gas cooling will also drive gas motions within the cluster cores. The bottom left panel in Figure~\ref{fig:cosmo} shows the the velocity dispersion and the velocity shear in the AGN run. The mode of the distribution of the velocity dispersion is about $400\;\kms$, which is increased by a factor of $\sim 2$ from the NR run. The distribution is also more spread out. This reduces the fraction of clusters having velocity dispersion consistent with the {\em Hitomi} observations, from $\sim 1/3$ to $\sim 1/7$.  The large increase in velocity dispersion is due to shock waves generated from the thermal blast in the AGN run. 

The velocity shear also increases in the AGN run. While the velocity shear distribution peaks at around $500\;\kms$, which is significantly larger than Perseus value, a significant fraction of clusters ($\sim 20\%$) have shear values consistent with Perseus.  The clusters that have velocity dispersion and shear matching those of the Perseus are more likely to be relaxed and disturbed respectively, although the dependence on dynamical state is quite weak in the AGN run. The turbulent-to-thermal pressure ratio increases slightly in the presence of AGN feedback, due to the increase in the velocity dispersion. 

The subgrid thermal AGN feedback implemented in our cosmological simulation does not resolve the AGN physics that is responsible for driving gas motions. Also, AGN feedback is likely to be kinetic instead of thermal, as suggested by jets and bubbles observed in the cores of real galaxy clusters. This shows that subgrid models based on thermal AGN feedback in cosmological simulations are likely not able to reproduce the level gas motions in real clusters. Because of this, coupled with the fact that thermal AGN feedback already fails to reproduce the thermal ICM properties in cluster cores, we choose not to pursue subgrid feedback models, in particular purely thermal ones, further. Instead, we focus next in Section~\ref{sec:isolated_sim} on studying the impact of ``gentle'' kinetic AGN feedback on core cluster gas motions by using high-resolution isolated AGN feedback simulations. 

\subsection{High-resolution Hydrodynamic Simulation with Self-Regulated Kinetic AGN feedback}
\label{sec:isolated_sim}

\begin{figure*}
\begin{center}
\subfigure{\includegraphics[scale=0.6]{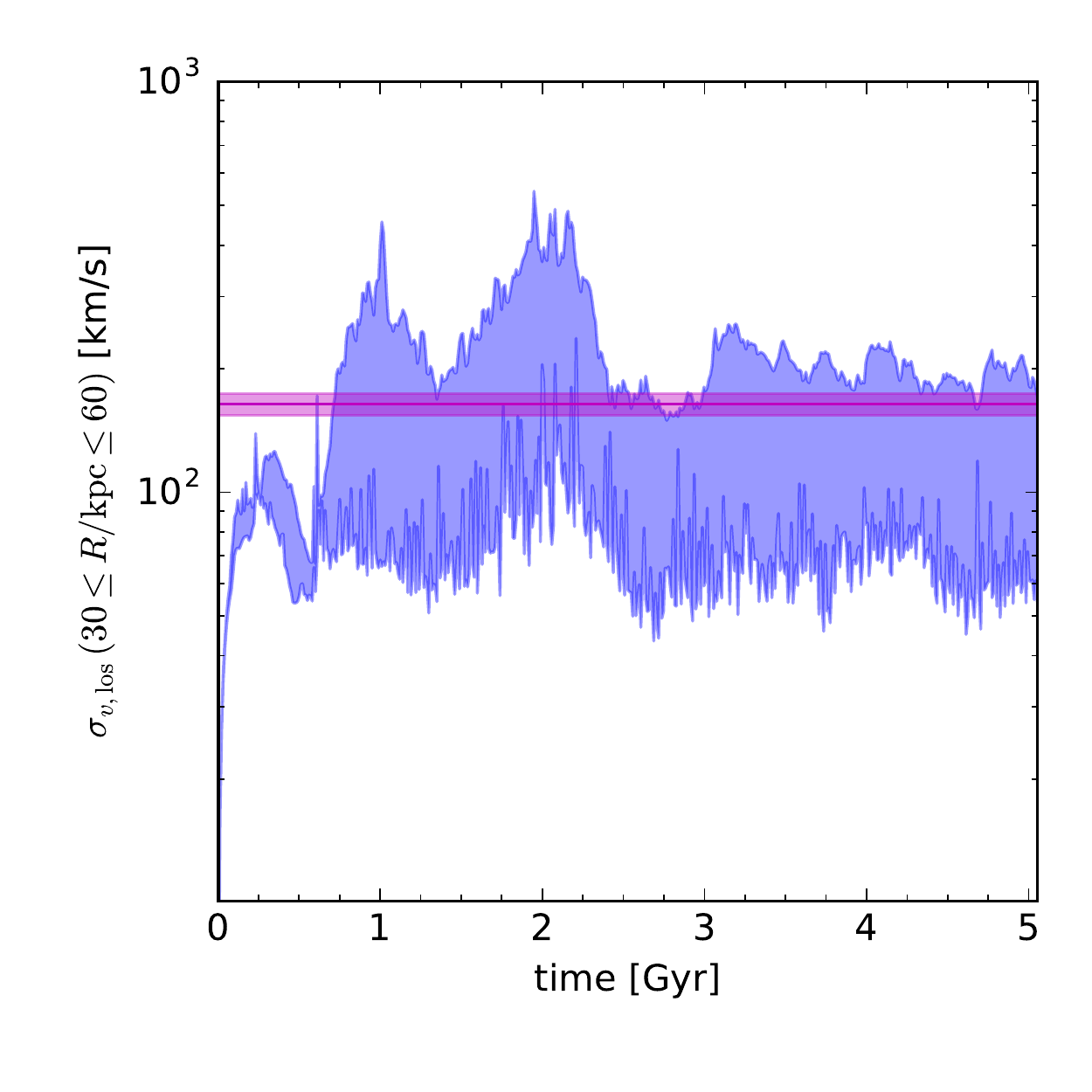}}
\subfigure{\includegraphics[scale=0.6]{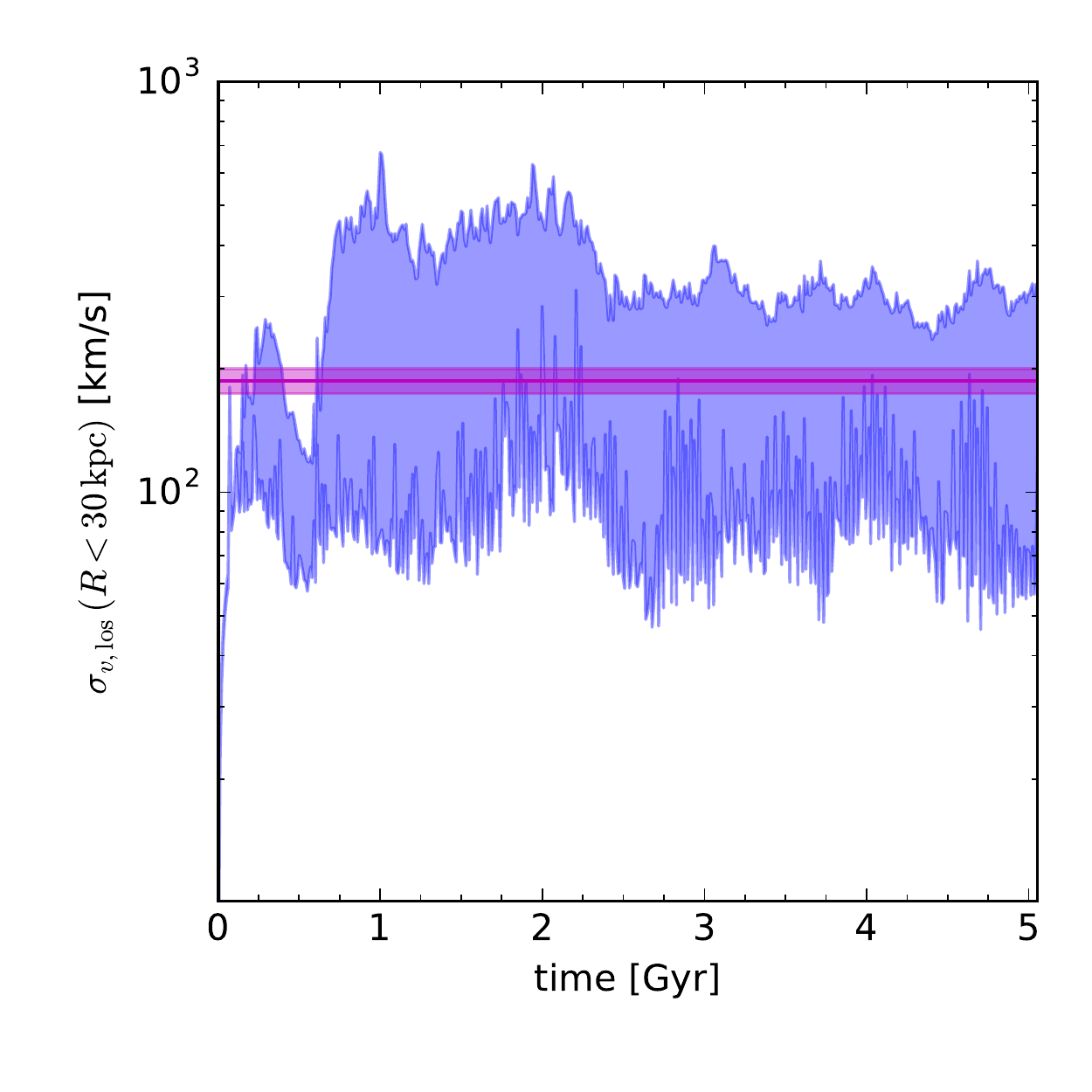}}
\subfigure{\includegraphics[scale=0.6]{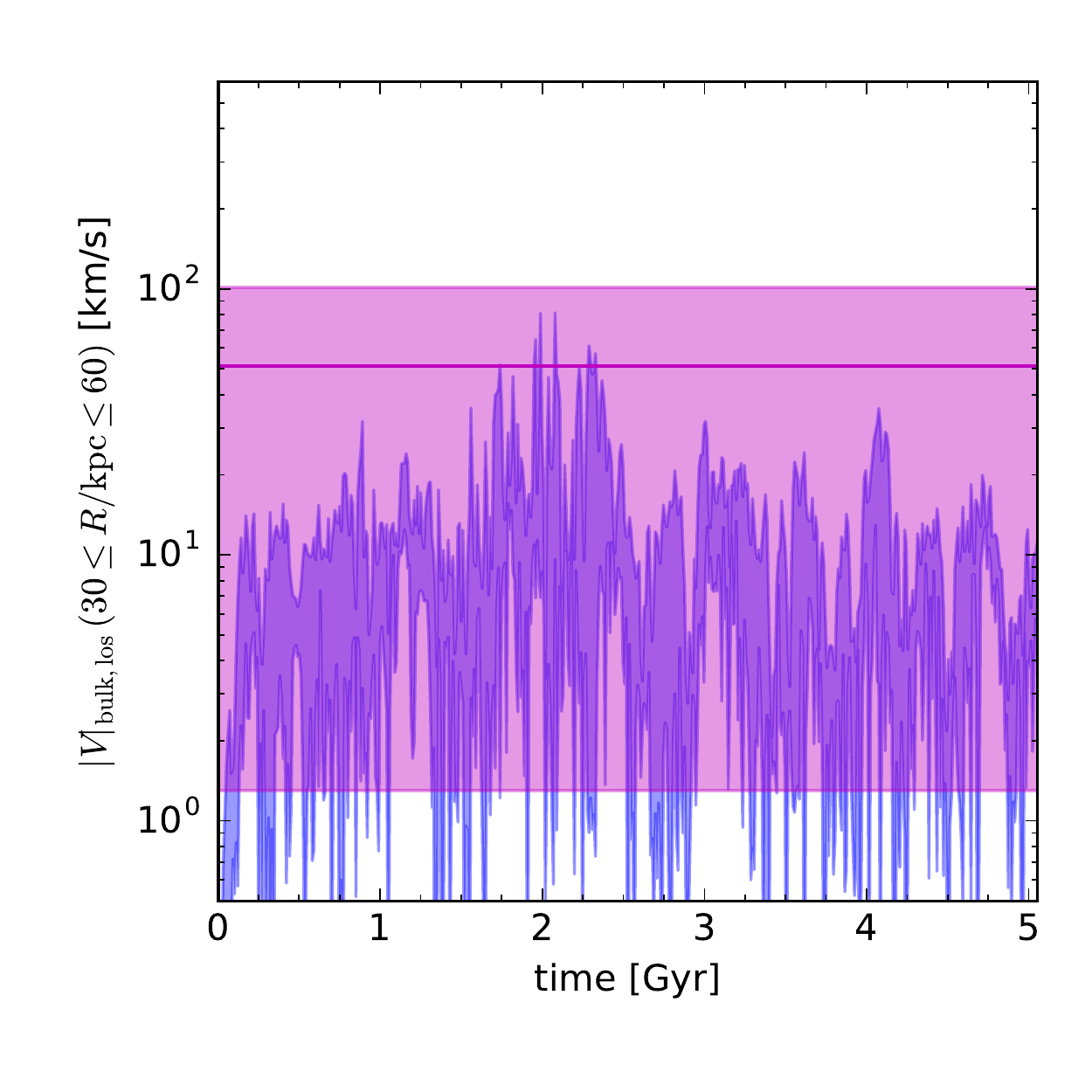}}
\subfigure{\includegraphics[scale=0.6]{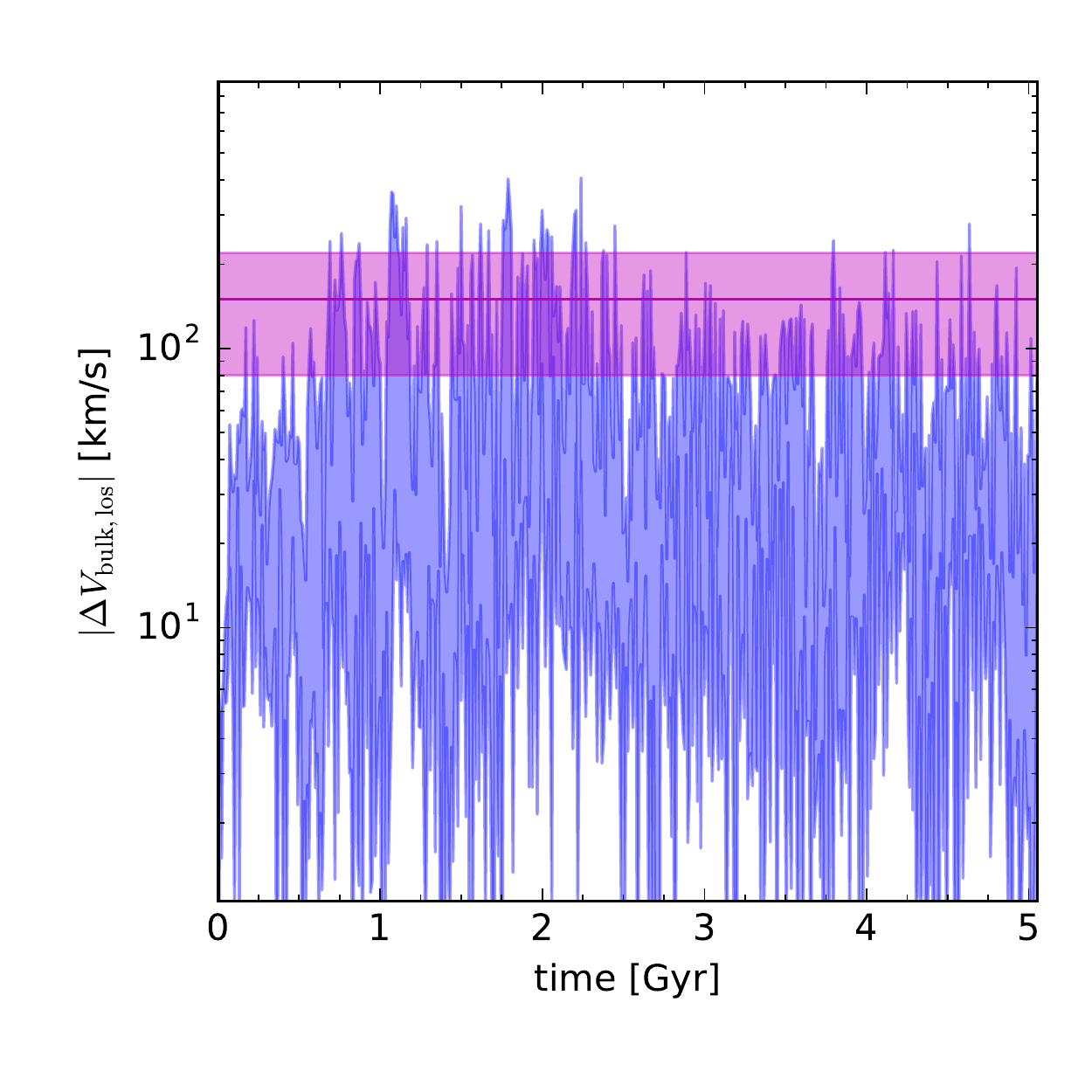}}
\caption
{High-resolution simulation with AGN outflow feedback self-regulated via chaotic cold accretion (CCA) in an isolated massive galaxy cluster: evolution for the \ion{Fe}{25}-\ion{Fe}{26} luminosity-weighted LOS velocity dispersion in the outer ({\em top left}) and inner ({\em top right}) projected annular bin, $R = [30,60]$~kpc and $R < 30$~kpc, respectively; while the {\em bottom left} and {\em bottom right} panels show the evolution of luminosity-weighted LOS bulk velocity and velocity shear, respectively. 
The velocity is measured along 3 random lines of sight at each time (507 snapshots with 10 Myr step), with the shaded blue area depicting the minimum to maximum range. The magenta band marks the {\em Hitomi} measurements with $1\sigma$ errorbar. AGN feedback drives significant turbulence up to $100$~kpc via frequent gentle outbursts, which drive velocity dispersions of a few $100\,\kms$, consistent with the {\em Hitomi} observation of Perseus, in particular during the fully developed CCA rain. AGN feedback drives mild bulk velocity and shear, which typically trace the large-scale motions, and is thus expected to be complemented by the cosmological accretion flows and mergers. Note, however, that the bulk velocity is almost unconstrained due to the large systematic calibration error ($\pm50\,\kms$).
}
\label{fig:G12}
\end{center}
\end{figure*}

We analyze the hydrodynamic simulation of self-regulated kinetic AGN feedback in an isolated massive galaxy cluster described in Section~\ref{sec:G12}.
Compared to the cosmological simulations, the spatial resolution in the core is 20 times higher, which allows us to resolve the AGN accretion and feedback physics, and the interaction between the AGN outflows/jets and the ICM. This simulation and the previous cosmological run should thus bracket the two complementary regimes responsible for generating turbulent and bulk motions in the ICM.

\subsubsection{AGN feedback: Velocity Dispersion} \label{sec:G12_sigma}

In the top left panel of Figure~\ref{fig:G12}, we show the \ion{Fe}{25}-\ion{Fe}{26}
X-ray luminosity-weighted velocity dispersion in the projected bin $R = [30,60]$~kpc. The velocity is computed along 3 random lines of sight every 10~Myr for the 5~Gyr evolution, with the shaded area covering the minimum to maximum range.

During the initial 0.7~Gyr, the relatively few AGN outflow events have not yet been able to stir the gas sufficiently to generate volume-filling turbulence. The initially imparted perturbations on top of the hydrostatic profiles help to sustain a mild level of ICM turbulence. However, the LOS velocity dispersion always remains below the {\em Hitomi} detection by a factor of 2 to 3.
This shows that a few single outbursts generating bubbles and shocks are not able to drive efficient turbulence \citep[e.g.,][]{reynolds_etal15}, although this is not the typical state of AGN feedback.

After roughly a Gyr, the situation changes. The cluster core has entered into full self-regulation. The CCA cascade is fully developed, promoting nonlinear multiphase condensation and intense cold/warm gas raining onto the central SMBH, boosting the accretion rate via inelastic collisions. The outflows become more frequent with variability over 2 orders of magnitude \citep[for flicker noise variability driven by CCA, see][]{Gaspari:2017_CCA}. The core of the cluster is now stirred more efficiently and the typical LOS velocity dispersions can easily reach the $165\,\kms$ detected by {\em Hitomi} at $R\sim 30-60$~kpc, with some reaching to $2.5$ times this value.
Within $30$~kpc (Figure~\ref{fig:G12}, top right panel), the velocity dispersion has relatively similar trend but is on average $\sim  50\%$ higher, consistent with the {\em Hitomi} detection of $187\pm13\,\kms$.

After a few Gyr, the CCA rain becomes less intense as the initial cosmic weather decays, resulting in a lower velocity dispersion that is still consistent with {\em Hitomi}. 
During this phase, the cluster has undergone $\sim 100$ outbursts, which have promoted a chaotic ICM atmosphere; 
SMBH accretion continues to be tightly self-regulated, preventing drastic oscillations in the cool-core structure for at least $5$~Gyr. Via such long-term simulated evolution, we thus probed all 3 major phases: the initially mild cosmic-weather dominated phase, the strong concurrent AGN feedback plus weather phase, and the third significant pure AGN feedback phase.

Overall, self-regulated kinetic AGN feedback can drive significant LOS velocity dispersion of $\sim 100-300\,\kms$ to at least $60$~kpc, in agreement with the {\em Hitomi} observations. We expect the velocity dispersion reaches its maximum after the system has entered full self-regulation with a fully developed cycle of CCA multiphase cascade and AGN outflows. Such a state is also consistent with the H$\alpha$ and molecular filamentary structures observed in Perseus core. 

\subsubsection{AGN feedback: Velocity Shear} \label{sec:vbulk}

While AGN feedback can drive significant velocity dispersion (linked to small-scale turbulent motions), the typical bulk velocity at $R = [30,60]$~kpc is around a few 10$\;\kms$, with maximum values up to $80\,\kms$ (Figure~\ref{fig:G12}, bottom left panel).
Unfortunately, the {\em Hitomi} detection of bulk motions is not well constrained as each pixel suffers a systematic calibration error of $\pm50\,\kms$ on top of a statistical error of $\pm20\,\kms$ (cf.~Figure~6 in H16). 
Since only one pixel reaches $94\,\kms$, it is safe to say that bulk velocities are much more contained than velocity dispersions, as found in the AGN feedback runs, at least within $60$~kpc.

The bottom right panel of Figure~\ref{fig:G12} shows the LOS bulk velocity gradient between the $30$~kpc and $60$~kpc, as done in the previous section. The magenta line shows the velocity gradient calculated from the southeast to northwest field of view in Perseus cluster, with an error band of $\pm70\,\kms$, albeit this is likely to be underestimated as only one pixel is really above such noise. 
Keeping in mind such uncertainties, kinetic AGN feedback only generates a mild shear.
The velocity shear varies greatly with time: $1/3$ of the time the central shear is comparable to the {\em Hitomi} estimates.

Overall, large-scale bulk and shear motions driven by AGN feedback within each feedback cycle are relatively mild, albeit still within current weak observation constraints. Thus we expect the AGN-driven shear to be complemented by shear driven on larger scales, such as cosmic accretion and mergers, as shown in Section~\ref{sec:cosmo_sim}.


\section{Discussion} 
\label{sec:disc}



\subsection{Implications for AGN feedback}

Despite evidence of ongoing AGN feedback activity in the Perseus Cluster, such as X-ray cavities and buoyant bubbles, the {\em Hitomi} observation do not show a high level of gas motions in the cluster core ($\lesssim 60$~kpc). However, this does not immediately imply that there is a strong tension between observations and models of AGN feedback. From what we know of radio-loud AGN, the injected AGN power in Perseus could plausibly reach $\sim 10^{45}$\,erg\,s$^{-1}$ with outflows in the nuclear region ($R\lesssim 100$~pc) up to $10^4\,\kms$. What {\em Hitomi} observes, however, is a luminosity-weighted projection along the full line of sight of the cluster over a much larger region ($R \lesssim 50$~kpc). Since AGN feedback is an inside-out process, the outflow loses power at progressively larger radii, as it entrains gas mass along the way. As demonstrated here, line-of-sight  averaged velocity dispersions $\sim 100\,\kms$ for the hot X-ray emitting gas at $50$~kpc can in fact be a natural outcome of AGN feedback -- as long as the feedback is ``gentle'' and well-mixed and leads to many small outbursts of AGN activity instead of a few isolated powerful ones.  

In order to contribute significantly to the velocity dispersions observed by {\em Hitomi}, the energy injected by the AGN should not thermalize quickly, remaining almost fully kinetic at least at the kpc scale, as shown in the G12 runs. Injecting mostly thermal (instead of kinetic) energy at sub-kpc scales would lead to a highly inefficient production of turbulence and central overheating via shocks (e.g., \citealt{Meece_etal16}).

With the high spectral resolution of {\em Hitomi}, any significiant velocity structures along the line-of-sight (such as those coming from AGN feedback) would have shown as large deviation in the \ion{Fe}{25} K$\alpha$ emission line profiles. The fact that no such deviation has been observed suggests that the ICM in the Perseus core is fairly well-mixed. In addition to the velocity constraints, the high-spectral resolution of {\em Hitomi} also allowed us to measure the thermal structure of the Perseus core along the LOS with unprecedented precision (Hitomi Collaboration, in prep.).  These observational constraints exclude any strong violent feedback events that create large spatial variations in gas properties. The gentle kinetic feedback scenario presented in this paper, in which feedback energy is injected frequently and in small quantities, allows efficient mixing of the feedback energy in the cluster core that produces velocity and thermal structures consistent with the {\em Hitomi} observations. This feedback model is also in line with the density and temperature profile of cool-core clusters observed in X-ray in the past decade \citep[e.g.,][]{McNamara:2016}. If the AGN feedback is instead bursty with low-frequency cycles \citep[e.g.,][]{Li_etal15}, it tends to overheat the cluster gas, raising the temperature higher than observed. The bursty feedback also means that substantial velocity dispersions are reached only during a few peaks (every $\sim$\,1\,-\,2 Gyr).
This would require that {\em Hitomi} measured the velocity dispersion during a rare AGN outburst phase in the Perseus cluster \citep[e.g.,][]{Li_etal16}, which seems to be too coincidental. 

In the self-regulated AGN feedback simulation presented in this paper, the AGN outflows are expected to be thermalized in the core via weak shocks, turbulent dissipation, and buoyant hot bubbles mixing with the ambient ICM \citep[see also][]{hillel_soker17}. It is difficult to disentangle the exact fraction participating in the different stages as they are highly interlinked \citep[e.g.,][]{Yang:2016}. Nevertheless, even if we assume that all gas motions go into turbulence dissipation, the related power per unit volume $P_{\rm turb} \approx \rho\sigma_{v,{\rm 3D}}^3/L_{\rm inj}\simeq 7.3\times 10^{-27}\,{\rm erg\,s^{-1}\,cm^{-3}}$ (for gas electron density $n_{\rm e}=0.01$\,cm$^{-3}$, injection scale $L_{\rm inj}=20$~kpc, and $\sigma_{v,{\rm 3D}}=286\,\kms$)
is only $\sim\,10\%$ of the typical AGN injection in the runs with $10^{45}$\,erg\,s$^{-1}$ deposited over a spherical volume with radius $50$~kpc. Nevertheless, turbulence can be an important component of the AGN feedback loop, in particular it allows the circulation of the feedback energy and the formation of ICM perturbations to trigger condensation of the observed multiphase gas. On the other hand, turbulence appears to be a complementary -- but not an exclusive -- process of heating the cluster core. Indeed, if turbulent dissipation (with timescale of $\sim 10$ Eddy turnover times, $\propto \mathrm{Mach^{-2}_{3D}}\,t_{\rm eddy}$) would completely heat the cluster core, turbulent diffusion (on much shorter timescale $\sim t_{\rm eddy}$) would have already washed out the entropy profile gradient, which is not observed in cool-core clusters like Perseus. 

At $R \gta 100$~kpc, AGN feedback is well complemented by the motions driven by cosmic accretion: infalling galaxies/groups, mergers, and filamentary accretion. Cosmic accretion-driven gas flows are likely to be an essential component in the AGN feedback loop. First, the cosmic-driven turbulence provides the initial perturbations in the density and temperature field of the ICM near the cluster cores, which helps promote gas condensation and accretion onto the SMBH. Second, large-scale bulk and shear motions can enhance the mixing and advection of AGN feedback energy and metals from the cluster center, heating and enriching the cluster core more uniformly and isotropically. 

\subsection{Limitations of the current work}

There are several caveats that must be kept in mind when interpreting our results. First, even after the initial effects of cosmic weather have died out in our simulations, the kinetic AGN feedback can still drive gas motions to the level of $100\;\kms$ close to the observed value (albeit in the upper envelope). Cosmological simulations {\em without} any AGN feedback, however, can also reach similar level of gas motions. For a kinetic AGN feedback model to work in a
more realistic cosmological context, where cosmic accretion continually
stirs the core (albeit at a low level), extra microphysics such as thermal conduction (e.g., \citealt{kannan_etal17}), magnetic field, viscosities, and plasma instabilities, may be required to either damp gas motions in the cluster cores, or to suppress some of gas motions from the AGN. 
At the same time, higher resolution is also required for the cosmological simulations to properly resolve the turbulence cascade in the cluster core.

Second, the cosmic weather implemented in the isolated simulation in this paper does not model large-scale shear generated from mergers and accretion, which might alter the BH accretion rate via tidal torques and thus the level of AGN-driven motions.
The complex, non-linear interactions between cosmic accretion, AGN feedback, and ICM microphysics will be addressed in future cosmological simulations that self-consistently model these processes. We are actively working on implementing the kinetic AGN feedback model into the cosmological runs, whose outputs will be used in upcoming papers to address the possible differences in the observational signatures (e.g., injection scales) between cosmic accretion and AGN feedback.  

Clearly, a larger observational sample beyond Perseus is also required.  With the eventual launch of the X-ray Astronomy Recovery Mission (XARM, the successor to {\em Hitomi}), {\em Athena}, and the addition of mm observations of cold gas feeding phase in the cluster core (e.g., by ALMA) we should be able build up such a sample and begin to dissect any correlations of the velocity dispersion and bulk motions with the feedback cycle and different core evolutionary periods, such as the pre-injection/feeding-dominated regime versus the post-injection/feedback-dominated regime. An X-ray calorimeter mission with high angular resolution, $\lesssim 1''$, such as for proposed for the {\em Lynx} mission, would be able to directly probe the small-scale velocity and thermal structures which are currently washed out in $\sim 1'$ resolution
{\em Hitomi} data. This should substantially improve the constraints we can place on the details of AGN feedback models. 

\subsection{Implications for hydrostatic mass estimates}

Lastly, it is also worth mentioning that the contribution from turbulent pressure support is expected to increase substantially towards the outskirts of clusters \citep[e.g.,][]{nelson_etal14b,Khatri:2016}, up to an order of magnitude  higher
than its value at the cluster core, due to the effects of cosmic accretion. While the turbulent pressure fraction measured at $60$~kpc by {\em Hitomi} is small, it does 
{\it not} imply that it is small at $R_{500}$ too, as larger radii entail larger injection scales and thus higher turbulent energies. Therefore, while the hydrostatic mass bias due to turbulence in the inner region may be small, the hydrostatic mass bias in the outer regions -- which is crucial for cluster cosmology applications -- can still be large, as corroborated by cosmological simulations.  


\section{Conclusions}
\label{sec:summary}


By using full cosmological simulations and high resolution hydro simulations of isolated galaxy clusters, we compared the theoretical estimates of gas motions in cluster cores with the first direct measurements of velocity dispersions in the core of Perseus cluster by \citet{hitomi16}. We summarize our findings below.

\begin{itemize}

\item 
In a cosmological representative sample of galaxy clusters from cosmological hydrodynamic simulations {\em without} baryonic physics (no radiative cooling, star formation, or subgrid feedback) where core gas motions are purely shaped by cosmic flows (mergers and accretion), one third of the mass-limited cluster sample have LOS velocity dispersion and shear levels consistent with the {\em Hitomi} observations within the projected radii of $30-60$~kpc. Such systems are mostly relaxed clusters which have not undergone major mergers in the past 4~Gyr. The cosmic-induced gas flows represent an {\em irreducible} background of gas motions.  Cosmological simulated clusters with thermal AGN feedback, on the other hand, over-predict the level of velocity dispersion and velocity shear. 

\item 
In the high-resolution (isolated) cluster simulation with cosmic-driven turbulence providing an initial Gaussian ICM perturbation field, kinetic AGN feedback is able to frequently drive LOS velocity dispersions of a few $100-300\;\kms$ within the projected radii of $30-60$~kpc (and $\sim 50\%$ higher in the central $30$~kpc), in agreement with the {\em Hitomi} observations. The velocity dispersion peaks during major chaotic cold accretion phases, which are augmented with cosmic gas flows. The bulk motions and velocity shear driven by the AGN alone are mild, with the latter reaching the observed {\em Hitomi} level one third of the time. 

\item 
Taken together, our results suggest that the cosmic accretion and kinetic AGN feedback are the key {\em complementary} drivers of gas motions in cluster cores.  While the gentle, self-regulated kinetic AGN feedback sustains significant velocity dispersions -- tracing the small-scale turbulent motions -- in the inner $\sim 10$~kpc. This effect of kinetic feedback diminishes rapidly with increasing radial distance. The large-scale velocity shear at $\gta 50$~kpc is instead sustained by continuous injection due to mergers, infalling groups, and penetrating streams. 
\end{itemize}

This work represents the first step in understanding the origins of gas motions in galaxy clusters. Future work should investigate the effects of AGN feedback on the velocity and thermodynamic structure of the ICM using high-resolution cluster simulations that properly capture both the microphysics of ICM (such as thermal conduction and viscosity, magnetic fields, and plasma instabilities) as well as cosmological gas flows (including the large-scale velocity field driven by mergers and streams). 

Observationally, it will be important to extend measurements to larger dynamic range as well as radial coverage to probe the full inertial range of the ICM turbulence \citep{zuhone_etal16}, and to study the effects of bulk and turbulent motions on the hydrostatic mass estimates of galaxy clusters \citep{lau_etal13}. It will also be important to increase the angular resolution beyond $\sim 1'$ angular resolution of {\em Hitomi} to probe gas flows closer to the cluster center for better constraints on the AGN feedback models. 
Future X-ray missions, such as XARM, {\em Athena}, and the proposed {\em Lynx} mission, will improve measurements of gas motions out to large radii, provide higher angular resolution, and look at a larger sample of clusters in order to probe the physical origins of their gas motions.\\

\acknowledgments 
This work is supported in part by NSF grant AST-1412768, NASA
ATP grant NNX11AE07G, NASA Chandra grants GO213004B and GO7-18121X, 
NASA ASTRO-H (Hitomi) Science Support grant NNX16AD55G, the
Research Corporation, and by the facilities and staff of the Yale Center for Research Computing. 
M.G. is supported by NASA through Einstein Postdoctoral Fellowship Award Number PF5-160137 issued by the Chandra X-ray Observatory Center, which is operated by the SAO for and on behalf of NASA under contract NAS8-03060.
HPC resources were in part provided by the NASA/Ames HEC Program (SMD-16-7251).
FLASH code was in part developed by the DOE NNSA-ASC OASCR Flash center at the University of Chicago.
Part of the analysis were performed using the publicly-available yt package \citep{turk_etal11}.

\bibliographystyle{yahapj}
\bibliography{references}

\end{document}